\documentclass{llncs}
\usepackage{a4wide}
\usepackage{graphicx}
\usepackage{amsmath,amssymb}

\newcommand{\isotope} [2] {\ensuremath{{}^{#2}\!{\mathrm{#1}}}}
\newcommand{\ket}     [1] {\ensuremath{\left|#1\right\rangle}}

\def\eps  {\varepsilon}
\def\up   {\uparrow}
\def\down {\downarrow}

\def\FF   {\ensuremath{\mathcal{F}}}

\def\KK   {\ensuremath{\mathcal{F}_k}}
\def\JJ   {\ensuremath{\mathcal{F}_{k-1}}}
\def\II   {\mathcal{I}}
\def\AA   {\mathcal{F}_\mathrm{All}}
\def\BB   {\ensuremath{\mathcal{B}}}

\usepackage{float}
\floatstyle{ruled}
\newfloat{algorithm}{h}{lop}
\floatname{algorithm}{Algorithm}
\newfloat{scheme}{h}{lop}
\floatname{scheme}{Scheme}
\newfloat{example}{h}{lop}
\floatname{example}{Example}

 \usepackage{latexsym}
 \newtheorem{thm}{Theorem}%[section] 
\title{Optimal Algorithmic Cooling of Spins}
\author{Yuval Elias$^1$, Jos{\'e} M. Fernandez$^2$, Tal Mor$^3$, and Yossi 
Weinstein$^4$}
\institute{ \small 1.
Chemistry Department, Technion, Haifa, Israel.
\\ \small 2.
D\'epartement de g\'enie informatique, \'Ecole Polytechnique de Montr\'eal, Montr\'eal, Canada
\\ \small 3.
Computer Science Department, Technion, Haifa, Israel.
\\ \small 4.
Physics Department, Technion, Haifa, Israel.}
\date{\today}

\begin{document}
\maketitle
\begin{abstract}
\emph{Algorithmic Cooling (AC) of Spins} is potentially the first 
near-future application of quantum computing devices. Straightforward quantum 
algorithms combined with novel entropy manipulations can result in a method to 
improve the identification of molecules. 

We introduce here several new exhaustive cooling algorithms, such as the 
Tribonacci and $k$-bonacci algorithms. In particular, we present the 
``all-bonacci'' algorithm, which appears to reach the maximal degree of cooling 
obtainable by the optimal AC approach.
\end{abstract}

\section{Introduction}
Molecules are built from atoms, and the nucleus inside each atom has a property 
called ``spin''. The spin can be understood as the orientation of the nucleus, 
and when put in a magnetic field, certain spins are binary, either up (ZERO) or 
down (ONE). Several such bits (inside a single molecule) represent a binary 
string, or a register. A macroscopic number of such registers/molecules can be 
manipulated in parallel, as is done, for instance, in Magnetic Resonance 
Imaging (MRI). The purposes of magnetic resonance methods include the 
identification of molecules (e.g., proteins), material analysis, and imaging, 
for chemical or biomedical applications. From the 
perspective of quantum computation, the spectrometric device that typically 
monitors and manipulates these bits/spins can be considered a simple ``quantum 
computing'' device. 

Enhancing the sensitivity of such methods is a Holy Grail in the area of 
Nuclear Magnetic Resonance (NMR). A common approach to this problem, known as 
``effective cooling'', has been to reduce the entropy of spins. A spin with 
lower entropy is considered ``cooler'' and provides a better signal when used 
for identifying molecules. To date, effective cooling methods have been plagued 
by various limitations and feasibility problems. 

``Algorithmic Cooling''~\cite{BMRVV02,FLMR04,AAC-pat} is a novel and 
unconventional effective-cooling method that vastly reduces spin entropy. 
AC makes use of ``data compression'' algorithms (that are run on the spins 
themselves) in combination with ``thermalization''. Due to Shannon's entropy 
bound (source-coding bound~\cite{CT91}), data compression alone is highly 
limited in its ability to reduce entropy: the total entropy of the spins in a 
molecule is preserved, and therefore cooling one spin is done at the expense 
of heating others. Entropy reduction is boosted dramatically by taking 
advantage of the phenomenon of thermalization, the natural return of a spin’s 
entropy to its thermal equilibrium value where any information encoded on the 
spin is erased. Our entropy manipulation steps are designed such that the 
excess entropy is always placed on pre-selected spins, called ``reset bits'', 
which return very quickly to thermal equilibrium. Alternating data compression 
steps with thermalization of the reset spins thus reduces the total entropy of 
the spins in the system far beyond Shannon's 
bound.%\footnote{See~\cite{FLMR04} for a simple explanation regarding Shannon's 
%entropy bound on cooling a single spin.} 
The AC of short molecules is 
experimentally feasible in conventional NMR labs; we, for example, recently 
cooled spins of a three-bit quantum computer beyond Shannon's entropy 
bound~\cite{POTENT}. 

\subsection{Spin-Temperature and NMR Sensitivity}
For two-state systems (e.g. binary spins) there is a simple 
connection between temperature, entropy, and probability. The 
difference in probability between the two states is called the 
\emph{polarization bias}. Consider a single spin particle in a constant magnetic
field. At equilibrium with a thermal heat-bath the probabilities of this spin 
to be up or down (i.e., parallel or anti-parallel to the magnetic field) are given by: 
$p_{\uparrow}=\frac{1+\eps_0}{2}$, and $p_{\downarrow}=\frac{1-\eps_0}{2}$.  
We refer to a spin as a bit, so that 
$\ket{\up}\equiv\ket{0}$ and $\ket{\down}\equiv \ket{1}$, where 
\ket{x} represents the spin-state $x$. The polarization bias
is given by $\eps_0=p_{\uparrow}-p_{\downarrow}=
\tanh \left(\frac{\hbar \gamma B}{2K_{B}T}\right)$, where $B$ is the magnetic 
field, $\gamma$ is the particle-dependent gyromagnetic constant,\footnote{This 
constant, $\gamma$, is thus responsible for the difference in the equilibrium 
polarization bias of different spins [e.g., a hydrogen nucleus is about 4~times
more polarized than a \isotope{C}{13} nucleus, but less polarized by three 
orders of magnitude than an electron spin].} $K_B$ is Boltzmann's coefficient, 
and $T$ is the thermal heat-bath temperature. Let 
$\eps=\frac{\hbar\gamma B}{2K_B T}$ such that $\eps_0=\tanh{\eps}.$ For high 
temperatures or small biases, higher powers of $\eps$ can be neglected, so we 
approximate $\eps_0\approx\eps.$ Typical values of $\eps_0$ for nuclear 
spins (at room temperature and a magnetic field of $\sim10$ Tesla) are 
$10^{-5}-10^{-6}.$

% 
% Parag: NMR - approaches to improve SNR - problems, limitations. 
%
A major challenge in the application of NMR techniques is to enhance
sensitivity by overcoming 
difficulties related to the Signal-to-Noise Ratio (SNR). 
Five fundamental approaches were traditionally suggested for improving the SNR
of NMR\@. Three straightforward approaches - cooling the entire system, 
increasing the magnetic field, and using a larger sample - are all expensive 
and limited in applicability, for instance they are incompatible with live 
samples. Furthermore, such approaches are often impractical due to sample or 
hardware limitations. A fourth approach - 
repeated sampling - is very feasible and is often employed in NMR experiments. 
However, an improvement of the SNR by a factor of M requires $M^2$ repetitions 
(each followed by a significant delay to allow relaxation), making this 
approach time-consuming and overly costly. Furthermore, it is inadequate for 
samples which evolve over the averaged time-scale, for slow-relaxing spins, or 
for non-Gaussian noise. 

\subsection{Effective cooling of spins}
The fifth fundamental approach to the SNR problem consists of cooling the 
spins without cooling the environment, an approach known as ``effective 
cooling'' of the spins~\cite{INEPT,Sorensen89,SV99,FHG+01}. The 
effectively-cooled spins can be used for spectroscopy until they relax to 
thermal equilibrium. The following calculations are done to leading order in 
$\eps_0$ and are appropriate for $\eps_0\ll1$. A spin temperature at 
equilibrium is $T\propto\eps_0^{-1}$. The single-spin Shannon entropy is
$H=1-\left(\eps_0^2/\ln 4 \right).$  A spin temperature out of thermal 
equilibrium is similarly defined (see for instance~\cite{Slichter90}). 
Therefore, increasing the polarization bias of a spin beyond its equilibrium 
value is equivalent to cooling the spin (without cooling the system) and to 
decreasing its entropy. 

Several more recent approaches are based on the creation of very
high polarizations, for example dynamic nuclear polarization~\cite{AFGH+03}, 
para-hydrogen in two-spin systems~\cite{ABCD+04}, and hyperpolarized 
xenon~\cite{OS04}. In addition, there are other spin-cooling methods, based on 
general unitary transformations~\cite{Sorensen89} and on 
(closely related) data compression methods in closed systems~\cite{SV99}. 

One method for effective cooling of spins, \emph{reversible polarization 
compression (RPC)}, is based on entropy manipulation techniques. RPC can be 
used to cool some spins while heating others~\cite{Sorensen89,SV99}. 
Contrary to conventional data compression,\footnote{Compression of 
data~\cite{CT91} such as bits in a computer file, can be performed by 
condensation of entropy to a minimal number of high entropy bits, which are 
then used as a compressed file.} RPC techniques focus on the low-entropy 
spins, namely those that get colder during the entropy manipulation process. 
RPC, also termed ``molecular-scale heat engine'', consists of reversible, 
in-place, lossless, adiabatic entropy manipulations in a closed 
system~\cite{SV99}. Therefore, RPC is limited by the second law of 
thermodynamics, which states that entropy in a closed system cannot decrease, 
as is also stated by Shannon's source coding theorem~\cite{CT91}. Consider the 
total entropy of $n$ uncorrelated spins with 
equal biases, $H(n) \approx n(1-\eps_0^2/\ln 4).$ This entropy could be 
compressed into $m \ge n(1-\eps^2/\ln 4)$ high entropy spins, leaving $n - m$ 
extremely cold spins with entropy near zero. Due to preservation 
of entropy, the number of extremely cold spins, $n - m$, cannot exceed
$ n\eps_0^2/\ln 4$. With a typical $\eps_0 \sim 10^{-5}$, extremely long 
molecules ($\sim 10^{10}$ atoms) are required in order to cool a single spin 
to a temperature near zero. If we use smaller molecules, with $n\ll 10^{10}$, 
and compress the entropy onto $n - 1$ fully-random spins, the entropy of the 
remaining spin satisfies~\cite{FLMR04} 

\begin{equation}  
1 - \eps^2_{\rm final} \ge n(1 - \eps_0^2/\ln4) - (n-1) = 1 - n\eps_0^2/\ln4. 
\end{equation}
Thus, the polarization bias of the cooled spin is bounded by
\begin{equation}\label{Eq:Shannon-bound} 
\eps_{\rm final}\le\eps_0\sqrt{n}\ . 
\end{equation}

When all operations are unitary, a stricter bound than imposed by entropy 
conservation was derived by S{\o}rensen~\cite{Sorensen89}. In practice, due to 
experimental limitations, such as the efficiency of the algorithm, relaxation 
times, and off-resonance effects, the obtained cooling is significantly below 
the bound given in eq~\ref{Eq:Shannon-bound}.

\begin{figure}[ht]
\begin{center}
\includegraphics[scale=0.3]{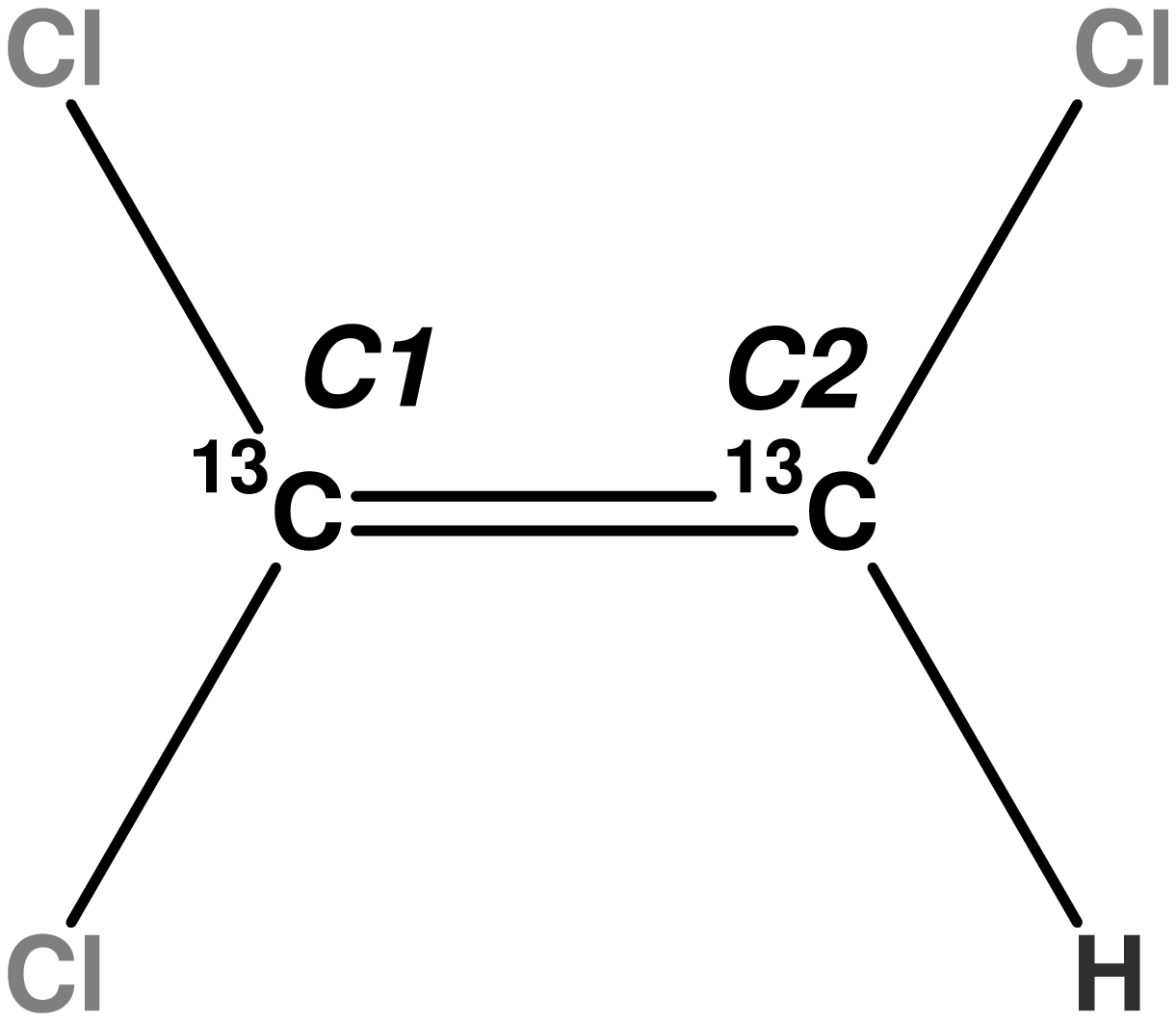}\end{center}
\caption{{\em A 3-bit computer: a TCE molecule labeled with two 
\isotope{C}{13}}. TCE has three spin nuclei: two carbons and a hydrogen,
which are named C1, C2 and H\@. The chlorines have a very small signal and their 
coupling with the carbons is averaged out. Therefore, TCE acts as a three-bit 
computer. The proton can be used as a reset spin because relative to the 
carbons, its equilibrium bias is four times greater, and its thermalization 
time is much shorter. Based on the theoretical ideas presented 
in~\cite{FLMR04}, the hydrogen of TCE was used to cool both carbons, decrease 
the total entropy of the molecule, and bypass Shannon's bound on cooling via 
RPC~\cite{POTENT}.}
\label{fig:TCE}
\end{figure}

Another effective cooling method is known as \emph{polarization transfer 
(PT)}\cite{INEPT,EP93}. This technique may be applied if at thermal equilibrium
the spins to be used for spectroscopy (the observed spins) are less polarized 
than nearby auxiliary spins. In this case, PT from the auxiliary spins to the 
observed spins is equivalent to cooling the observed spins (while heating the 
auxiliary spins). PT from one spin to another is limited as a cooling 
technique, because the polarization bias increase of the observed spin is 
bounded by the bias of the highly polarized spin. PT is regularly used in NMR 
spectroscopy, among nuclear spins on the same molecule~\cite{INEPT}. As a 
simple example, consider the 3-bit molecule trichloroethylene (TCE) shown in 
Fig.~\ref{fig:TCE}.
 The hydrogen nucleus is about four times more polarized than each of the 
carbon nuclei; PT from a hydrogen can be used to cool a single carbon by a 
factor of four. A different form of PT involves shifting entropy from nuclear 
spins to electron spins. This technique is still under 
development~\cite{FHG+01,OS04}, but has significant potential in the future. 

%Fig.~\ref{fig:compare-methods}a illustrates the limitation of RPC techniques 
%which arises from thermodynamic considerations (see eq~\ref{Eq:Shannon-bound} 
%above).
% \begin{figure}[ht]
%\begin{center}
%\includegraphics[scale=0.4]{PAC-RPC-grey}\end{center}
%\caption{{\em A comparison of the space complexities of RPC and PAC\@.}
%The number of spins required for a 5-fold improvement in the polarization bias
%of a single spin by RPC, PAC1 and PAC2, when all spins have the same initial
%polarization bias $\eps_0$.
%Colors illustrate the temperature: grey representing thermal equilibrium,
%blue and red indicating colder and hotter, respectively.
%\underline{Case A (RPC)}:
%the number of required spins is calculated from
%Shannon's bound~(Eq.~\ref{Eq:Shannon-bound}) --- 25 spins are
%required [Note that any specific RPC algorithm will generally perform worse
%than this theoretical bound, requiring more spins~\cite{Sorensen89,SV99}].
%\underline{Case B (PAC1)}:
%the algorithm ``PAC1'' (Practicable Algorithmic
%Cooling 1) requires 18 spins.
%%YW-11May: removed ref to appendix
%%
%% (see Equation~\ref{Eq:PAC1-space}).
%Note that with one additional (parallel) PT step and one
%additional RESET step all hot spins regain their initial bias.
%\underline{Case C (PAC2)}:
%the algorithm ``PAC2'' (Practicable Algorithmic Cooling 2) --- 9 spins are 
%required.[A remark: A 25-fold improvement in polarization via RPC, PAC1 or 
%PAC2 requires 625 spins, 34 spins and 17 spins, respectively.]}
%\label{fig:compare-methods}
%\end{figure}
Unfortunately, the manipulation 
of many spins, say $n > 100$, is a very difficult task, and the gain of 
$\sqrt{n}$ in polarization is not substantial enough to justify putting this 
technique into practice. 

In its most general form, RPC is applied to spins with different initial 
polarization biases, thus PT is a special case of RPC\@. We sometimes refer to 
both techniques and their combination as \emph{reversible algorithmic cooling}. 

\section{Algorithmic Cooling} \label{sec:AC}
Boykin, Mor, Roychowdhury, Vatan, and Vrijen (hereinafter referred to as BMRVV),
coined the term \emph{Algorithmic Cooling (AC)} for their novel 
effective-cooling method~\cite{BMRVV02}. AC expands previous effective-cooling 
techniques by exploiting entropy manipulations in \emph{open systems}. It 
combines RPC with relaxation (namely, thermalization) of the 
\emph{hotter spins}, in order to cool far beyond Shannon's entropy bound.

AC employs slow-relaxing spins (which we call \emph{computation spins}) and 
rapidly relaxing spins (\emph{reset spins}), to cool the system by pumping 
entropy to the environment. Scheme~1 details the three basic 
operations of AC\@. The ratio $R_{relax-times}$, between the spin-lattice 
relaxation times of the computation spins and the reset spins, must satisfy 
$R_{relax-times}\gg 1$, to permit the application of many cooling steps to 
the system.

In all the algorithms presented below, we assume that the relaxation time of 
each computation spin is sufficiently large, so that the entire algorithm is 
completed before the computation spins lose their polarization.

The practicable algorithmic cooling (PAC) suggested in~\cite{FLMR04} indicated 
a potential for near-future application to NMR spectroscopy~\cite{AAC-pat}. 
In particular, it presented an algorithm (named PAC2) which uses any odd number
of spins such that one of them is a reset spin and the other $2L$ spins are 
computation spins. PAC2 cools the spins such that the coldest one can 
(ideally) reach a bias of $(3/2)^L$. This proves an exponential advantage of 
AC over the best possible reversible algorithmic cooling, as reversible 
cooling techniques (e.g., of refs~\cite{Sorensen89} and~\cite{SV99}) are 
limited to a bias improvement factor of $\sqrt{n}$. As PAC2 can be applied 
to small $L$ (and small $n$), it is potentially suitable for near future 
applications.
\\
\textbf{Scheme~1: }
\emph{AC is based on the combination of three distinct operations:
 \begin{enumerate}
  \item RPC\@. Reversible Polarization Compression steps redistribute the 
   entropy in the system so that some computation spins are cooled while other 
   computation spins become hotter than the environment.
  \item SWAP\@. Controlled interactions allow the hotter computation spins to 
   adiabatically lose their entropy to a set of reset spins, via PT from the 
   reset spins onto these specific computation spins.
  \item WAIT\@. The reset spins rapidly thermalize, conveying their entropy to 
   the environment, while the computation spins remain colder, so that the 
   entire system is cooled.
 \end{enumerate}
% \label{sch:AC-op}
}

\subsection{Block-Wise Algorithmic Cooling} \label{sec:AC1}
The original Algorithmic Cooling (BMRVV AC)~\cite{BMRVV02} was designed to 
address the scaling problem of NMR quantum computing. Thus, a significant 
number of spins are cooled to a desired level and arranged in a consecutive 
block, such that the entire block may be viewed as a register of cooled spins. 
The calculation of the cooling degree attained by the algorithm was based on 
the law of large numbers, yielding an exponential reduction in spin 
temperature. Relatively long molecules were required due to this statistical 
nature, in order to ensure the cooling of 20, 50, or more spins (the algorithm 
was not analyzed for a small number of spins). All computation spins were 
assumed to be arranged in a linear chain where each computation spin (e.g., 
\isotope{C}{13}) is attached to a reset spin (e.g., \isotope{H}{1}).
%, see Fig.~\ref{fig:compare-methods}b.
 Reset and computation spins are assumed to have the same bias $\eps_0$. BMRVV 
AC consists of applying a recursive algorithm repeating (as many times as 
necessary) the sequence of RPC, SWAP (with reset spins), and WAIT, as detailed 
in Scheme~1. 
%YW-11May: removed refence to appendices.
%
%For a basic description of logic gates, see 
%Appendix~\ref{app:Logic-Gates}. 
The relation between gates and NMR pulse sequences is
discussed in ref~\cite{EFMW-IJC}.

This cooling algorithm employed a simple form of RPC termed 
\emph{Basic Compression Subroutine (BCS)}~\cite{BMRVV02}. The computation spins
are ordered in pairs, and the following operations are applied to each pair of computation spins $X,Y$. $X_j,Y_j$ denote the state of the respective spin 
after stage $j$; $X_0,Y_0$ indicate the initial state.
\begin{enumerate}
 \item Controlled-NOT (CNOT), with spin $Y$ as the control and spin $X$ as 
  target: $Y_1 = Y_0, X_1 = X_0\oplus Y_0$, where $\oplus$ denotes exclusive 
  OR (namely, addition modulo~2). This means that the target spin is flipped 
  (NOT gate – $\ket{\up}\leftrightarrow\ket{\down}$) if the control spin is 
  \ket{1} (i.e., \ket{\down}). Note that $X_1 = \ket{0}\Leftrightarrow 
  X_0 = Y_0$, in which case the bias of spin $Y$ is doubled, namely the 
  probability that $Y_1 = \ket{0}$ is very close to $(1+2\eps_0)/2.$ 
 \item NOT gate on spin $X_1$. So $X_2 = NOT(X_1)$. 
 \item Controlled-SWAP (CSWAP), with spin X as a control: if $X_2 = \ket{1}$  
 ($Y$ was cooled), transfer the improved state of $Y$ (doubled bias) to a 
 chosen location by alternate SWAP and CSWAP gates.
\end{enumerate}
Let us show that indeed in step 1 the bias of spin $Y$ is doubled whenever 
$X_1 = \ket{0}$. Consider the truth table for the CNOT operation below: 
$$
\begin{array}{cccc}
input : &   X_0  Y_0  &    output : &  X_1  Y_1 \\
        &   0\:\: 0  & \rightarrow &  0\:\: 0  \\
        &   0\:\: 1  & \rightarrow &  1\:\: 1  \\
        &   1\:\: 0  & \rightarrow &  1\:\: 0  \\
        &   1\:\: 1  & \rightarrow &  0\:\: 1  
\end{array}
$$
The probability that both spins are initially \ket{0} is 
$p_{00}\equiv P\left(X_0=\ket{0},Y_0=\ket{0}\right)=(1+\eps_0)^2/4$. 
In general, $p_{kl}\equiv P\left(X_0=\ket{k},Y_0=\ket{l}\right)$ such that 
$p_{01} = p_{10} = (1+\eps_0)(1-\eps_0)/4$, and $p_{11} = (1-\eps_0)^2/4$. 
After CNOT, the conditional probability $P\left(Y_1=\ket{0}|X_1=\ket{0}\right)$
is $\frac{q_{00}}{q_{00}+q_{01}}=\frac{p_{00}}{p_{00}+p_{11}}$, where 
$q_{kl}\equiv P\left(X_1=\ket{k},Y_1=\ket{l}\right)$ and $q_{kl}$ is derived 
from $p_{kl}$ according to the truth table, so that $q_{00} = p_{00}$, and 
$q_{01} = p_{11}$. 
%YW-11May: removed reference to appendix
%
%See also Appendix~\ref{app:Logic-Gates}. 
In terms of $\eps_0$ this probability is 
$$
\frac{\left(1+\eps_0\right)^2}{\left(1+\eps_0\right)^2+\left(1-\eps_0\right)^2}
\approx\frac{1+2\eps_0}{2},
$$
indicating that the bias of $Y$ was indeed doubled in this case.

For further details regarding the basic compression subroutine and its 
application to many pairs of spins (in a recursive manner) to reach a bias of 
$2^j\eps_0$, we refer the reader to ref~\cite{BMRVV02}.

\subsection{Practicable Algorithmic Cooling (PAC)}
An efficient and experimentally feasible AC technique was later presented, 
termed \emph{``practicable algorithmic cooling (PAC)''}~\cite{FLMR04}. Unlike 
the first algorithm, the analysis of PAC does not rely on the law of large 
numbers, and no statistical requirement is invoked. PAC is thus a simple 
algorithm that may be conveniently analyzed, and which is already applicable 
for molecules containing very few spins. Therefore, PAC has already led to 
experimental implementations. PAC algorithms use PT steps, reset steps, and 
\emph{3-bit-compression (3B-Comp)}. As already mentioned, one of the algorithms 
presented in~\cite{FLMR04}, PAC2, cools the spins such that the coldest one can
(ideally) reach a bias of $(3/2)^L$, while the number of spins is only $2L+1.$ 
PAC is simple, as all compressions are applied to three spins, often with 
identical biases. The algorithms we present in the next section are more 
efficient and lead to a better degree of cooling, but are also more complicated.
We believe that PAC is the best candidate for near future applications of AC, 
such as improving the SNR of biomedical applications. 

PAC algorithms~\cite{FLMR04} use a basic 3-spin RPC step termed 
3-bit-compression (3B-Comp):
\\
\textbf{Scheme~2: }\emph{3-Bit­Compression (3B­Comp)
 \begin{enumerate}
   \item CNOT, with spin $B$ as a control and spin $A$ as a target.  Spin $A$ is
         flipped if $B = \ket{1}$. 
   \item NOT on spin $A$.
   \item CSWAP with spin $A$ as a control and spins $B$ and $C$ as targets. $B$
         and $C$ are swapped if $A = \ket{1}$.
 \end{enumerate}
%\label{sch:3BC}
}
Assume that the initial bias of the spins is $\eps_0$. The result of 
scheme~2 is that spin $C$ is cooled: if $A = \ket{1}$ after the 
first step (and $A = \ket{0}$ after the second step), $C$ is left unchanged 
(with its original bias $\eps_0$); if however, $A = \ket{0}$ after 
the first step (hence $A = \ket{1}$ after the second step), spin $B$ is cooled 
by a factor of about~2 (see previous subsection), and following the CSWAP the 
new bias is placed on $C$. Therefore, on average, $C$ is cooled by a factor 
of~3/2. We do not care about the biases of the other two spins, as they 
subsequently undergo a reset operation.

In many realistic cases the polarization bias of the reset spins at thermal 
equilibrium, $\eps_0$, is higher than the biases of the computation spins. Thus,
an initial PT from reset spins to computation spins (e.g., from hydrogen to 
carbon or nitrogen), cools the computation spins to the $0^{\mathrm{th}}$ 
purification level, $\eps_0.$

As an alternative to Scheme~2, the 3B-Comp operation depicted in 
Scheme~3 is similar to the CNOT-CSWAP combination 
(Scheme~2) and cools spin $C$ to the same degree. This gate is 
known as the MAJORITY gate since the resulting value of bit $C$ indicates 
whether the majority of the bits had values of \ket{0} or \ket{1} prior to the 
operation of the gate.
\\
\textbf{Scheme 3: }\emph{Single operation implementing 3B-Comp\\
%\begin{eqnarray} 
Exchange the states $\ket{100}\leftrightarrow\ket{011}$.\\ 
Leave the rest of the states unchanged.
%{\rm input:CBA} \ &  \quad \quad {\rm output:CBA} \nonumber \\ 
%\quad \quad 000 \ & \rightarrow \quad \quad \quad 000   \nonumber \\ 
%\quad \quad 001 \ & \rightarrow \quad \quad \quad 001   \nonumber \\ 
%\quad \quad 010 \ & \rightarrow \quad \quad \quad 010   \nonumber \\ 
%\quad \quad 011 \ & \rightarrow \quad \quad \quad 100   \nonumber \\ 
%\quad \quad 100 \ & \rightarrow \quad \quad \quad 011   \nonumber \\ 
%\quad \quad 101 \ & \rightarrow \quad \quad \quad 101   \nonumber \\ 
%\quad \quad 110 \ & \rightarrow \quad \quad \quad 110   \nonumber \\ 
%\quad \quad 111 \ & \rightarrow \quad \quad \quad 111.  \nonumber 
%\end{eqnarray}
%\label{sch:3BC-1op}
}\\
If 3B-Comp is applied to three spins \{C,B,A\} which have identical biases,
$\eps_C=\eps_B=\eps_A=\eps_0$, then spin $C$ will acquire a new bias $\eps'_C.$
This new bias is obtained from the initial probability that spin $C$ is \ket{0}
  by adding the effect of exchanging $\ket{100}\leftrightarrow\ket{011}$: 
\begin{eqnarray}
\frac{1+\eps'_C}{2}&=&\frac{1+\eps_C}{2}+p_{\ket{100}}-p_{\ket{011}}\\
\nonumber  &=&\frac{1+\eps_0}{2}+\frac{1-\eps_0}{2}\frac{1+\eps_0}{2}
  \frac{1+\eps_0}{2}-\frac{1+\eps_0}{2}\frac{1-\eps_0}{2}\frac{1-\eps_0}{2}
         =\frac{1+\frac{3\eps_0-\eps_0^3}{2}}{2}.
\end{eqnarray}
The resulting bias is
\begin{equation}\label{eq:3B-Comp-exact-identicalbiases}
\eps'_C = \frac{3\eps_0- \eps_0^3}{2}
\ ,
\end{equation}
and in the case where $\eps_0\ll 1$
\begin{equation}\label{eq:3B-Comp-identicalbiases}
\eps'_C  \approx\frac{3\eps_0}{2}.
\end{equation}

% 
% Parag: 3B-Comp - the two methods we have seen
% 
We have reviewed two schemes for cooling spin $C$: one using CNOT and CSWAP 
gates (see Scheme~2), and the other using the MAJORITY gate (see 
Scheme~3). 
Spin $C$ reaches the same final bias following both schemes. The other spins may
obtain different biases, but this is irrelevant for our purpose, as they 
undergo a reset operation in the next stages of any cooling algorithm. 
%YW-9May: the section ``Experimental Algorithmic Cooling'' is commented out
%
%NMR implementations of various gates and protocols are described in section 
%``Experimental Algorithmic Cooling''.

\begin{figure}[ht]
\begin{center}
\includegraphics[scale=0.25]{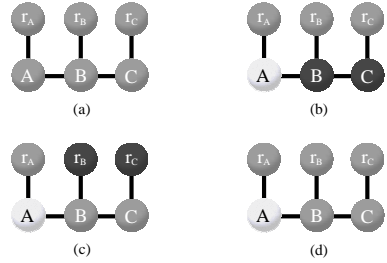}\end{center}
\caption{
An abstract example of a molecule with three computation spins $A$, $B$ and $C$,
attached to reset spins $r_A$, $r_B$, and $r_C$, respectively.
All spins have the same equilibrium polarization bias, $\eps_0$ (a).
The temperature after each step is illustrated by color:
gray - thermal equilibrium;
white - colder than initial temperature;
and black - hotter than initial temperature.
PAC uses 3-bit compression (3B-Comp),
Polarization Transfer (PT) steps, and RESET steps:
1. 3B-Comp($C;B;A$); the outcome of this step is shown in (b).
2. PT($r_B \rightarrow B$), PT($r_C \rightarrow C$); the outcome is shown in (c).
3. RESET($r_B,r_C$); the outcome is shown in (d).
The 3-bit-compression applied in the first step operates on the
three computation spins, increasing the bias of spin $A$ by a factor of $3/2$,
while heating the other two spins. The 3B-Comp step cools spin $A$, and the 
following PT and RESET steps restore the initial biases of the other spins, 
thus the entire system is cooled.
}
\label{fig:ABCmolecule}
\end{figure}

The simplest practicable cooling algorithm, termed Practicable Algorithmic 
Cooling 1 (PAC1)~\cite{FLMR04}, employs dedicated reset spins, which are not 
involved in compression steps. PAC1 on three computation spins is shown in 
Fig.~\ref{fig:ABCmolecule}, where each computation spin $X$ has a neighboring 
reset spin, $r_X$, with which it can be swapped. The following examples 
illustrate cooling by PAC1. In order to cool a single spin (say, spin $C$) to 
the first purification level, start with three computation spins, $CBA$, and 
perform the sequence presented in Example~1.
\footnote{This sequence may be implemented by the application of an appropriate sequence of radiofrequency pulses.
%; see ``Experimental Algorithmic Cooling''
} 
\\
\textbf{Example~1: }\emph{Cooling spin $C$ to the $1^\mathrm{st}$ purification 
level by PAC1
\begin{enumerate}
  \item PT($(r_C\rightarrow C);(r_B\rightarrow B);(r_A\rightarrow A)$), to initiate all spins.
  \item 3B-Comp($C;B;A$), increase the polarization of $C$.
\end{enumerate}
%\label{ex:PAC1-on-3-spins}
}

A similar algorithm can also cool the entire molecule beyond Shannon's entropy
bound. See the sequence presented in
Example~2.
\\
\textbf{Example~2: }\emph{Cooling spin $C$ and bypassing Shannon's entropy bound
 by PAC1
\begin{enumerate}
  \item PT($(r_C\rightarrow C);(r_B\rightarrow B);(r_A\rightarrow A)$), to initiate all spins.
  \item 3B-Comp($C;B;A$), increase the polarization of $C$.
  \item WAIT
  \item PT($(r_B\rightarrow B);(r_A\rightarrow A)$), to reset spins $A,B$.
\end{enumerate}
%\label{ex:PAC1-bypass-Shannon}
}
In order to cool one spin (say, spin $E$) to the second purification level 
(polarization bias $\eps_2$), start with five computation spins ($EDCBA$) and 
perform the sequence presented in Example~3. 

For small biases, the polarization of spin $E$ following the sequence in 
Example~3 is $\eps_2 \approx (3/2)\eps_1 \approx (3/2)^2 \eps_0$.
\\
\textbf{Example~3: }\emph{Cooling spin $E$ in $EDCBA$ to the $2\mathrm{nd}$ 
purification level by PAC1
\begin{enumerate}
  \item PT($E;D;C$), to initiate spins $EDC$.
  \item 3B-Comp($E;D;C$), increase the polarization of $E$ to $\eps_{1}$.
  \item WAIT
  \item PT($D;C;B$), to initiate spins $DCB$.
  \item 3B-Comp($D;C;B$), increase the polarization of $D$ to $\eps_{1}$.
  \item WAIT
  \item PT($C;B;A$), to initiate spins $CBA$.
  \item 3B-Comp($C;B;A$), increase the polarization of $C$ to $\eps_{1}$.
  \item 3B-Comp($E;D;C$), increase the polarization of $E$ to $\eps_{2}$.
\end{enumerate}
%\label{exa:PAC1-2}
}
For molecules with more spins, a higher cooling level can be obtained. 
%YW-11May: removed ref to appendix
%
%see Appendix~\ref{app:AC-complexity}. 

This simple practicable cooling algorithm (PAC1) is easily generalized to cool 
one spin to any purification level $L$.~\cite{FLMR04} The resultant bias will 
be very close to $(3/2)^L,$ as long as this bias is much smaller than~1. For a 
final bias that approaches~1 (close to a pure state), as required for 
conventional (non-ensemble) quantum computing, a more precise calculation is 
required.

Consider an array of $n$ computation spins,  $c_{n}c_{n-1}\ldots c_{2}c_{1}$, 
where each computation spin, $c_i$, is attached to a reset spin, $r_i$ 
(see Fig.~\ref{fig:ABCmolecule} for the case of $n=3$).
To cool $c_k$, the spin at index $k$, to a purification level 
$j \in {1 \ldots L}$ the procedure $M_j(k)$ was recursively defined as 
follows~\cite{FLMR04}: $M_0(k)$ is defined as a single PT step from reset spin 
$r_k$ to computation spin $c_k$ to yield a polarization bias of $\eps_0$ (the 
$0^\mathrm{th}$ purification level). The procedure $M_1(k)$ applies $M_0$ to the
three spins followed by 3B-Comp on these spins, so that spin $c_k$ is cooled to 
the first purification level. Similarly, $M_2(k)$ applies $M_1$ three times to 
cool $c_k;c_{k-1};c_{k-2}$ to the first purification level, followed by 3B-Comp 
on these spins, so that spin $c_k$ is cooled to the second purification level. 
We use the notation $\BB_{(j-1)\rightarrow j}(k)$ to represent the application 
of 3B-Comp to spins to purify spin $c_k$ from $\eps_{j-1}$ to $\eps_j$. Then, 
the full algorithm has a simple recursive form, described in 
Algorithm~1.  
\\
\textbf{Algorithm~1: }\emph{
Practicable algorithmic cooling 1 (PAC1):\\
%label{alg:PAC}
 For $j\in \{1,\ldots ,L\}$
\begin{equation}
M_{j}(k)={\mathcal{B}}_{\{(j-1)\longrightarrow j\}}(k)
M_{j-1}(k-2)\; M_{j-1}(k-1)\; M_{j-1}(k) \ ,
\end{equation}
applied from right to left ($M_{j-1}(k)$ is applied first).}
\\
For instance,$M_{1}(3)=\BB_{\{0\rightarrow 1\}}(3) M_{0}(1)\; M_{0}(2)\; 
M_{0}(3)$, is 3B-Comp applied after reset as described in 
Example~1.\footnote{The procedure of cooling one spin to the second level (starting with 
five spins) is written as $M_{2}(5)=\BB_{\{1\rightarrow 2\}}(5) M_{1}(3)\; 
M_{1}(4)\; M_{1}(5)$.}
Clearly, $M_1(k)$ can be applied to any $k \ge 3$,
$M_{2}(k)$ to $k \ge 5$, and $M_{j}(k)$ to $k \ge 2j+1$. Thus, to cool a single 
spin to a purification level of $L,$ $2L+1$ computation spins and an equal 
number of reset spins are required. A single reset spin could be used for 
initializing all relevant computation spins, at the expense of additional time 
steps.

Reset spins may also be used for compression, thus replacing the 3B-Comp and PT 
steps above by a generalized RPC\@. The corresponding algorithm, termed PAC2, has 
an improved space complexity relative to PAC1. We explicitly show how this is 
achieved. Let $\eps_0$ be the polarization bias of the reset spin. In order to 
cool a single spin to $\eps_1$, start with two computation spins, $CB$, and one 
reset spin, $A$, and perform the sequence shown in 
Example~4 to cool spin $C$.
\\
\textbf{Example~4: }\emph{Cooling spin $C$ to the $1^\mathrm{st}$ purification 
level by PAC2
\begin{enumerate}
  \item PT($A\rightarrow B$).
  \item PT($B\rightarrow C$) to initiate spin $C$.
  \item RESET($A$) (by waiting).
  \item PT($A\rightarrow B$) to initiate spin $B$.
  \item RESET($A$).
If the thermalization time of the computation spins is sufficiently
large, there are now three spins with polarization bias $\eps_0$.
  \item 3B-Comp to increase the polarization of spin $C$ to $\eps_1$.
\end{enumerate}
%\label{ex:PAC2-on-3-spins}
}
In order to cool one spin (say, spin $E$) to the second purification level
(polarization bias $\eps_{2}$), start with $5$ computation spins $(EDCBA)$ and
follow Example~5.
\\
\textbf{Example~5: }\emph{Cooling spin $E$ in EDCBA to the $2^\mathrm{nd}$ 
purification level by PAC2
\begin{enumerate}
  \item PT sequentially to initiate spins $EDC$ (RESET($A$) after each PT).
  \item 3B-Comp on spins $EDC$ to increase the polarization of spin $E$ to $\eps_1$.
  \item PT sequentially to initiate spins $DCB$ (RESET($A$) after each PT).
  \item 3B-Comp on spins $DCB$ to increase the polarization of spin $D$ to $\eps_1$.
  \item PT sequentially to initiate spins $CB$ (RESET($A$) after each PT).
  \item 3B-Comp on spins $CBA$ to increase the polarization of spin $C$ to $\eps_1$.
  \item 3B-Comp on spins $EDC$ to increase the polarization of spin $E$ to $\eps_2$.
\end{enumerate}
%\label{ex:PAC2-on-5-spins}
}
By repeated application of PAC2 in a recursive manner (as for PAC1),
spin systems can be cooled to very low temperatures. PAC1 uses dedicated reset 
spins, while PAC2 also employs reset spins for compression. The simplest 
algorithmic cooling can thus be obtained with as few as~3 spins, comprising~2 
computation spins and one reset spin.

%
% Parag: 3BComp on spins with different biases.
%
%YW-11May: removed ref to appendix
%
%In Appendix~\ref{app:AC-complexity} we provide an analysis of the space and and time complexity of 
%PAC1 and PAC2. 
The algorithms presented so far applied compression steps 
(3B-Comp) to three identical biases ($\eps_0$); recall that this cools one spin
 to a new bias, $\eps'_C\approx(3/2)\eps_0.$ Now consider applying compression to three spins with different biases 
$(\eps_C,\eps_B,\eps_A)$; spin $C$ will acquire a new bias $\eps'_C$, which is 
a function of the three initial biases~\cite{Jose-PhD-Thesis,Thesis}. This new 
bias is obtained from the initial probability that spin $C$ is \ket{0} by 
adding the effect of the exchange $\ket{100}\leftrightarrow\ket{011}$:
\begin{eqnarray}
&&\frac{1+\eps'_C}{2}=\frac{1+\eps_C}{2}+p_{100}-p_{011}\nonumber\\
&&=\frac{1+\eps_C}{2}+\frac{1-\eps_C}{2}\frac{1+\eps_B}{2}\frac{1+\eps_A}{2}
  -\frac{1+\eps_C}{2}\frac{1-\eps_B}{2}\frac{1-\eps_A}{2}\nonumber\\
&&=\frac{1+\frac{\eps_C+\eps_B+\eps_A-\eps_C\eps_B\eps_A}{2}}{2}.
\label{eq:3B-Comp-exact-probability}
\end{eqnarray}
The resulting bias is 
\begin{equation}
\label{eq:3B-Comp-exact}
\eps'_C = \frac{\eps_C + \eps_B + \eps_A - \eps_C\eps_B\eps_A}{2}
\ ,
\end{equation}
and in the case where $\eps_C,\eps_B,\eps_A\ll 1,$
\begin{equation}\label{eq:3B-Comp}
\eps'_C  \approx\frac{\eps_C+\eps_B+\eps_A}{2}\ .
\end{equation}

\section{Exhaustive Cooling Algorithms}
The following examples and derivations are to leading order in the biases. This 
is justified as long as all biases are much smaller than~1, including the final 
biases. For example, with $\eps_0 \sim 10^{-5}$ ($\eps_0$ is in the order of 
magnitude of $10^{-5}$) and $n \le 13$, the calculations are fine (see 
section~\ref{sec:optimal} for details).

\subsection{Exhaustive Cooling on Three Spins}
\textbf{Example~6: }\emph{Fernandez~\cite{Jose-PhD-Thesis}: 
$\mathcal{F}(C,B,A,m)$\\
 Repeat the following $m$ times
 \begin{enumerate}
   \item 3B-Comp($C;B;A$), to cool $C$.
   \item RESET($B;A$)
 \end{enumerate}
% \label{ex:Fernandez}
}
Consider an application of the primitive algorithm in Example~6 to three spins, $C,B,A$, where $A$ and $B$ are reset spins with initial biases of and (the index over the bias of $C$ denotes the iteration). After each WAIT 
step, spins $A$ and $B$ are reset back to their equilibrium bias, $\eps_0$.
As $B$ and $A$ play roles of both reset and computation spins, the operation 
RESET in Example~6 simply means WAIT\@. From eq~\ref{eq:3B-Comp}, after the first iteration
$$
\eps_C^{(1)}=\frac{\eps_C^{(0)}+\eps_B+\eps_A}{2}= \frac{0+2\eps_0}{2} = \eps_0.
$$
After the second iteration
$$
\eps_C^{(2)}=\frac{\eps_C^{(1)}+\eps_B+\eps_A}{2}=\frac{\eps_0+2\eps_0}{2}
=\frac{3\eps_0}{2}.
$$
After the $m^{\mathrm{th}}$ iteration
\begin{equation}
\eps^{(m)}_{C} =
        \frac{\eps^{(m-1)}_C+2\eps_0}{2} =
         2^{-m}\eps_C^{(0)}+2\eps_0\sum_{j=1}^{m}2^{-j}=
         0+\left(1-2^{-m}\right)2\eps_0.
\end{equation}
The asymptotic bias $(m \rightarrow \infty)$ may be extracted from 
\begin{equation}\label{eq:Fernandez-approximated-limit}
\eps_C^{(m)}\approx \frac{\eps_C^{(m)}+2\eps_0}{2},
\end{equation}
with the unique solution and bias configuration
\begin{equation}\label{eq:Fernandez-asymptotic-result}
\eps_C^{(m)}=2\eps_0\Rightarrow\{2\eps_0,\eps_0,\eps_0\}.
\end{equation}
In order to achieve good asymptotics, one would like to reach 
$\eps_C = (2-\delta)\eps_0$, where $\delta$ is arbitrarily small. In this 
case the number of iterations required is given by
$2^{1-m}=
\delta\Longrightarrow m =1+\left\lceil \log_2 (1/\delta)\right\rceil$.
For example, if $\delta= 10^{-5}$, 18~repetitions are sufficient. Up to an 
accuracy of $\delta$, the biases after the last reset are as in 
eq~\ref{eq:Fernandez-asymptotic-result}.

\subsection{The Fibonacci Algorithm}
An algorithm based on 3B-Comp was recently devised~\cite{SMW07}, which produces 
a bias configuration that asymptotically approaches the Fibonacci series. In 
particular, when applied to $n$ spins, the coldest spin attains the bias 
$\eps_{\mathrm{final}} \approx \eps_0 F_n$, where $F_n$ is the 
$n^{\mathrm{th}}$ element of the series and $\eps_0 F_n \ll1.$ 
Note that $\eps_C^{(m)}$ of eq~\ref{eq:Fernandez-asymptotic-result} is 
$\eps_0 F_3$. Also note that for 12~spins and $\eps_0 \sim 10^{-5}$, 
$\eps_0 F_{12} \ll1$, so the approximation is useful for non-trivial spin 
systems. Compare the bias enhancement factor in this case, $F_{12} = 144$, to 
PAC2 with 13~spins - $(3/2)^6\approx 11.$ 

Example~7 expands Example~6 to four spins with
initial biases $\eps_D = \eps_C = 0$ and $\eps_B = \eps_A = \eps_0$ ($A,B$ are 
reset spins). We refer to the parameter $m$ from Example~6 as 
$m_3$, the number of repetitions applied to three spins. We refer to $\FF$ from 
Example~6 as $\FF_2$ for consistency with the rest of the 
section. Consider the following example:
\\
\textbf{Example~7: }\emph{$ \FF_2(D, C, B, A, m_4, m_3)$\\
%  \label{ex:Fibonacci}
   Repeat the following $m_4$ times:
  \begin{enumerate}
     \item 3B-Comp($D;C;B$), places the new bias on spin $D$.
     \item $\FF_2(C, B, A, m_3)$.
  \end{enumerate}
}
After each iteration, $i$, the bias configuration is 
$\{\eps_D^{(i)},\eps_C^{m_3},\eps_0,\eps_0\}.$ Running the two steps in 
Example~7 exhaustively $(m_4, m_3 \gg1)$ yields: 
\begin{eqnarray}
\eps_C^{(m_{3})}&\approx& 2\eps_0, \nonumber\\
\eps_D^{(m_4)}&\approx& \frac{\eps_B+\eps_C^{(m_{3})}+\eps_D^{(m_4)}}{2} 
\label{eq:Fibonacci-on-4-spins}\\
&\Rightarrow& \eps_D^{(m_4)}= \eps_B+\eps_C^{(m_{3})}= 3\eps_0 = \eps_0F_4.\nonumber
\end{eqnarray}
This unique solution was obtained by following the logic of 
eqs~\ref{eq:Fernandez-approximated-limit} and~\ref{eq:Fernandez-asymptotic-result}.

We generalize this algorithm to $n$ spins $A_n,\ldots,A_1$ in 
Algorithm~2.
\\
\textbf{Algorithm~2: }
\emph{Fibonacci $\FF_2(A_n,\ldots,A_1, m_n,,m_3)$\\
%  \label{alg:Fibonacci-on-n-spins}
  Repeat the following $m_n$ times:
  \begin{enumerate}
    \item 3B-Comp$(A_n;A_{n-1};A_{n-2})$.
    \item $\FF_2(A_{n-1},\ldots,A_1,m_{n-1},\ldots,m_3).$
  \end{enumerate}
  [with $\FF_2(A_3,A_2,A_1,m_3)$ defined by Example~6.]
}\\
Note that different values of $m_{n-1},\ldots, m_3$ may be applied at each 
repetition of step 2 in Algorithm~2. This is a 
recursive algorithm; it calls itself with one less spin. Running 
Algorithm~2 exhaustively 
$(m_n, m_{n-1},\ldots,m_3 \gg  1)$ results, similarly to 
eq~\ref{eq:Fibonacci-on-4-spins}, in
\begin{eqnarray}
\eps_{A_n}^{(m_{n})}&\approx& \frac{\eps_{A_n}^{(m_{n})}+\eps_{A_{n-1}}+\eps_{A_{n-2}}}{2}
\nonumber\\
&\Rightarrow& \eps_{A_n}^{(m_n)}\approx \eps_{A_{n-1}}+\eps_{A_{n-2}}.
\end{eqnarray}
This formula yields the Fibonacci series $\{\ldots,8,5,3,2,1,1\},$ therefore 
$\eps_{A_i} \rightarrow\eps_0 F_i$. We next devise generalized algorithms 
which achieve better cooling. An analysis of the time requirements of the 
Fibonacci cooling algorithm is provided in~\cite{SMW07}. 

\subsection{The Tribonacci Algorithm}\label{sub:Tribonacci}
Consider 4-bit-compression (4B-Comp) which consists of an exchange between 
the states $\ket{1000}$ and \ket{0111} (the other states remain invariant 
similar to 3B-Comp). Application of 4B-Comp to four spins $D,C,B,A$ with 
corresponding biases $\eps_D,\eps_C,\eps_B,\eps_A \ll 1$ results in a 
spin with the probability of the state $\ket{0}$ given by 
\begin{equation}
\frac{1+\eps_D'}{2}=\frac{1+\eps_D}{2}+p_{\ket{1000}}-p_{\ket{0111}}
\approx\frac{1+\frac{\eps_A+\eps_B+\eps_C+3\eps_D}{4}}{2},\label{eq:4BC-exact}
\end{equation}
following the logic of eqs~\ref{eq:3B-Comp-exact-probability} 
and~\ref{eq:3B-Comp-exact}, and finally
\begin{equation}
\label{eq:4BC-result}
\eps_D'\approx (\eps_A+\eps_B+\eps_C+3\eps_D)/4.
\end{equation}
Example~8 applies an algorithm based on 4B-Comp to 4~spins, with 
initial biases $\eps_D = \eps_C = 0$ ($A, B$ are reset spins). 
In every iteration of Example~8, running step~2 exhaustively 
yields the following biases: $\eps_C = 2\eps_0, \eps_B = \eps_A = \eps_0$. 
The compression step (step~1) is then applied onto the configuration 
\\
\textbf{Example~8: }\emph{$\FF_3(D,C, B, A, m_4, m_3)$\\
  Repeat the following $m_4$ times:
  \begin{enumerate}
     \item \label{enu:4BC} 4B-Comp(D;C;B;A).
     \item \label{enu:Tribonacci-on-3-spins}$\FF_2(C,B,A,m_3)$.
  \end{enumerate}
%  \label{ex:F4_4}
}
{}From eq~\ref{eq:4BC-result}, $\eps^{(i+1)}_D=(4\eps_0+3\eps^{(i)}_D)/4$. For 
sufficiently large $m_4$ and $m_3$ the algorithm produces final polarizations of
\begin{equation} 
\eps^{(m_4)}_D \approx \eps_0+ \frac{3}{4}\eps^{(m_4)}_D\Longrightarrow  
\eps^{(m_4)}_D \approx 4\eps_0.
\end{equation}
For more than 4~spins, a similar 4B-Comp based algorithm may be defined. 
Example~9 applies an algorithm based on 4B-Comp to 5~spins.
\\
\textbf{Example~9: }\emph{$\FF_3(E,D,C,B,A,m_5,m_4,m_3)$\\
%\label{ex:F4_5}
  Repeat the following $m_5$ times:
  \begin{enumerate}
    \item 4B-Comp($E;D;C;B$).
    \item \label{enu:Tribonacci-on-4-spins}$\FF_3(D,C,B,A,m_4,m_3)$.
  \end{enumerate}
}
Step 2 is run exhaustively in each iteration; the biases of $DCBA$ after this
 step are $\eps_D = 4\eps_0, \eps_C  = 2\eps_0, B = A = \eps_0$.
The 4B-Comp step is then applied to the biases of spins $(E,D,C,B)$. 
Similarly to eq~\ref{eq:4BC-result}, 
$\eps^{(i+1)}_E=(7\eps_0+3\eps^{(i)}_E)/4.$ Hence, for sufficiently large $m_5$ 
and $m_4$ the final bias of $E$ is
\begin{equation}
\eps_E^{(m_5)}\approx (7\eps_0 + 3\eps^{(m_5)}_E)/4 
\Rightarrow \eps_E^{(m_5)}\approx 7\eps_0.
\end{equation} 
Algorithm~3 applies to an arbitrary number of spins $n > 4$. This 
is a recursive algorithm, which calls itself with one less spin.
% 
% Parag: Tribonacci on n spins 
% 
\\
\textbf{Algorithm~3: }
\emph{Tribonacci: $\FF_3(A_n,\ldots,A_1,m_{n},\ldots,m_3)$\\
Repeat the following $m_{n}$ times:
\begin{enumerate}
\item 4B-Comp($A_n;A_{n-1};A_{n-2};A_{n-3}$).
\item \label{enu:FF3-on-4-spins}$\FF_3(A_{n-1},
      \ldots,A_1,m_{n-1},\ldots,m_{3})$.
\end{enumerate}
[With $\FF_3(A_4,A_3,A_2,A_1,m_4,m_3)$ given by Example~8.]
%\label{alg:F3_n}
}\\
% 
% Parag: The resulting Tribonacci series. 
% 
The compression step, 4B-Comp, is applied to
$$\eps_{A_{n}}^{(i)},\eps_{A_{n-1}},\eps_{A_{n-2}},\eps_{A_{n-3}},$$
and results in
\begin{equation}
\eps_{A_n}^{(i+1)}=
(\eps_{A_{n-1}}+\eps_{A_{n-2}}+\eps_{A_{n-3}}+3\eps_{A_{n}}^{(i)})/4.
\end{equation}
For sufficiently large $m_{j},j=3,\ldots,n$
\begin{eqnarray}
\eps_{A_n}^{(m_{n})}&\approx& \nonumber
(\eps_{A_{n-1}}+\eps_{A_{n-2}}+\eps_{A_{n-3}}+3\eps_{A_{n}}^{(m_{n})})/4\\
&\Rightarrow& \eps_{A_n}^{(m_{n})}\approx \eps_{A_{n-1}}+\eps_{A_{n-2}}+\eps_{A_{n-3}}.
\end{eqnarray}
For $\eps_{A_3}=2\eps_0,\eps_{A_2}=\eps_{A_1}=\eps_0$, the resulting bias will
be $\eps_{A_n}^{(m_{n})}\approx\eps_0T_n$, where $T_n$ is the 
$n^{\mathrm{th}}$ Tribonacci number.\footnote{The Tribonacci series (also 
known as the Fibonacci 3-step series) is generated by the recursive formula 
$a_i = a_{i-1} + a_{i-2} + a_{i-3}$,  where $a_3 = 2$ and 
$a_2 = a_1 = 1.$} As for the Fibonacci algorithm, we assume
 $\eps_0 T_n \ll1$. The resulting series is $\{\ldots,24,13,7,4,2,1,1\}$. 

% 
% Parag: The k-bonacci series. 
% 
\subsection{The $k$-bonacci Algorithm}\label{sub:k-bonacci}
A direct generalization of the Fibonacci and Tribonacci algorithms above is 
achieved by the application of $(k+1)$-bit-compression, or $(k+1)$B-Comp. 
This compression on $k+1$ spins involves the exchange of \ket{100\cdots 000} 
and \ket{011\cdots 111}, leaving the other states unchanged. When 
$(k+1)$B-Comp is applied to $k+1$ spins with biases 
$\{\eps_{A_{k+1}},\eps_{A_{k}},\ldots,\eps_{A_2},\eps_{A_1}\},$  where $A_1$ 
and $A_2$ are reset spins, the probability that the leftmost spin is \ket{0} 
becomes (similarly to eq~\ref{eq:4BC-exact})
\begin{equation}
\frac{1+\eps_{k+1}'}{2}=\frac{1+\eps_{k+1}}{2}
       +p_{100\cdots 000}-p_{011\cdots 111}\approx
    \frac{1+\frac{(2^{k-1}-1)\eps_{A_{k+1}}
           +\sum_{j=1}^{k}\eps_{A_{j}}}{2^{k-1}}}{2}.
\end{equation}
Therefore, the bias of the leftmost spin becomes
\begin{equation}\label{eq:(k+1)BC-result}
\eps_{k+1}'\approx
\frac{(2^{k-1}-1)\eps_{A_{k+1}}+\sum_{j=1}^{k}\eps_{A_{j}}}{2^{k-1}}.
\end{equation}
Example~10 applies an algorithm based on 
$(k+1)$B-Comp to $k+1$ spins. $\FF_k$ on $k+1$ spins calls (recursively) 
$\FF_k-1$ on $k$ spins (recall that $F_3$ on four spins called $F_2$ on three
 spins).
\\
\textbf{Example~10: }\emph{$\KK(A_{k+1},\ldots,A_1,m_{k+1},\ldots,m_3)$\\
%  \label{ex:kbonacci-on-k+1-spins}
  repeat the following $m_{k+1}$ times:
  \begin{enumerate}
    \item $(k+1)$B-Comp($A_{k+1},A_{k},\ldots,A_2,A_1$)
    \item \label{enu:(k-1)bonacci-on-(k-1)-spins}
    $\JJ(A_{k},\ldots,A_1,m_k,\ldots,m_3)$
  \end{enumerate}
}
If step~\ref{enu:(k-1)bonacci-on-(k-1)-spins} is run exhaustively 
at each iteration, the resulting biases are 
$$
\eps_{A_{k}}=2^{k-2}\eps_0,\eps_{A_{k-1}}=2^{k-3}\eps_0,\ldots,
\eps_{A_3}=2\eps_0,\eps_{A_2}=\eps_{A_1}=\eps_0.
$$
The $(k+1)$B-Comp step is then applied to the biases 
$\eps_{A_{k+1}}^{(i)},2^{k-2}\eps_0,2^{k-3}\eps_0,\ldots
,2\eps_0,\eps_0,\eps_0.$
{}From eq~\ref{eq:(k+1)BC-result},
$$
\eps_{A_{k+1}}^{(i+1)}\approx
 \frac{(2^{k-1}-1)\eps_{A_{k+1}}^{(i)}+2^{k-1}\eps_0}{2^{k-1}}.
$$
Hence, for sufficiently large $m_j, j = 3,\ldots,k+1$, the final bias of 
$A_{k+1}$ is
\begin{equation}
\eps_{A_{k+1}}^{(m_{k+1})}
\approx \frac{(2^{k-1}-1)\eps_{A_{k+1}}^{(m_{k+1})}+2^{k-1}\eps_0}{2^{k-1}}
\Rightarrow \eps_{A_{k+1}}^{(m_{k+1})}\approx 2^{k-1}\eps_0.
\end{equation}
For more than $k+1$ spins a similar $(k+1)$B-Comp based algorithm may be 
defined. Example~11 applies such an algorithm to 
$k+2$ spins. 
\\
\textbf{Example~11: }\emph{$\KK(A_{k+2},\ldots,A_1,m_{k+2},\ldots,m_3)$\\
%  \label{ex:kbonacci-on-k+2-spins}
  Repeat the following steps $m_{k+2}$ times:
  \begin{enumerate}
    \item $(k+1)$B-Comp($A_{k+2},A_{k+1},\ldots,A_3,A_2$)
    \item \label{enu:kbonacci-on-k+1-spins}
       $\KK(A_{k+1},\ldots,A_1,m_{k+1},\ldots,m_3)$ [defined in Example~11.]
  \end{enumerate}
}
When step~\ref{enu:kbonacci-on-k+1-spins} is run exhaustively at each 
iteration, the resulting biases are 
$$
\eps_{A_{k+1}}=2^{k-1}\eps_0,\eps_{A_{k}}=2^{k-2}\eps_0,\ldots,\eps_{A_3}=2\eps_0,\eps_{A_2}=\eps_{A_1}=\eps_0.
$$
The $(k+1)$B-Comp is applied to the biases.
{}From Eq.~\ref{eq:(k+1)BC-result},
$$
\eps_{A_{k+2}}^{(i+1)}=
\frac{(2^{k-1}-1)\eps_{A_{k+2}}^{(i)}+(2^{k}-1)\eps_0}{2^{k-1}}.
$$

Hence, for sufficiently large $m_j, j=3,\ldots,k+2$, the final bias of $A_{k+2}$
 is
\begin{equation}
\eps_{A_{k+2}}^{(m_{k+2})}\approx
\frac{(2^{k-1}-1)\eps_{A_{k+2}}^{(m_{k+2})}+(2^{k}-1)\eps_0}{2^{k-1}}
\Rightarrow \eps_{A_{k+2}}^{(m_{k+2})}\approx (2^{k}-1)\eps_0.
\end{equation}
Algorithm~4 generalizes Examples~10
and~11.
\\
\textbf{Algorithm~4: }\emph{$k$-bonacci: $\KK(A_n,\ldots,A_1,m_n,\ldots,m_3)$\\
Repeat the following $m_{n}$ times:
\begin{enumerate}
%\label{alg:kbon}
\item $(k+1)$B-Comp($A_n,A_{n-1},\ldots,A_{n-k}$).
\item $\KK(A_{n-1},\ldots,A_1,m_{n-1},\ldots,m_3).$
\end{enumerate}
[with $\KK(A_{k+1},\ldots,A_1,m_{k+1},\ldots,m_3)$ defined in
Example~10.]
}\\
The algorithm is recursive; it calls itself with one less spin. The 
compression step, $(k+1)$B-Comp, is applied to 
$$\eps_{A_n}^{(i)},\eps_{A_{n-1}},\ldots\eps_{A_{n-k}}.$$ {}From
Eq.~\ref{eq:(k+1)BC-result}, the compression results in
\begin{equation}
\eps_{A_n}^{(i+1)}
\approx\frac{(2^{k-1}-1)\eps_{A_{n}}^{(i)}+\sum_{j=1}^{k}\eps_{A_{n-j}}}{2^{k-1}}.
\end{equation}
For sufficiently large $m_{n}, m_{n-1}$, etc.
\begin{eqnarray}
\eps_{A_n}^{(m_{n})}&\approx&
\frac{(2^{k-1}-1)\eps_{A_{n}}^{(m_{n})}+\sum_{j=1}^{k}\eps_{A_{n-j}}}{2^{k-1}}
\nonumber\\
&\Rightarrow& \eps_{A_n}^{(m_{n})}\approx \sum_{j=1}^{k}\eps_{A_{n-j}}
\end{eqnarray}
This set of biases corresponds to the $k$-step Fibonacci
sequence which is generated by a recursive formula.
\begin{equation} 
a_1,a_2=1,
a_\ell=
\left\{\begin{array}{ll}
\sum^{\ell-1}_{i=1}a_{\ell-i},& 3\le \ell \le k+1\\
~&\\
\sum^{k}_{i=1}a_{\ell-i} & \ell > k+1
\end{array}
\right\}.
\end{equation}
Notice that for $3\le \ell\leq k+1$,
\begin{equation}\label{eq:k-step-Fibonacci-sequence}
a_\ell=\sum_{i=1}^{\ell-1}a_i=
1+1+2+4+\cdots + 2^{\ell-4}+2^{\ell-3}=2^{\ell-2}.
\end{equation}
The algorithm uses $\ell$-bit-compression ($\ell$-B-Comp) gates, where 
$3\le \ell \le k+1$.

% 
% Parag: all-bonacci 
% 
\subsection{The All-bonacci Algorithm}\label{sub:all-bonacci}
In Example~11 ($\KK$ applied to $k+1$ spins), the 
resulting biases of the computation spins,
$2^{k-1}\eps_0,2^{k-2}\eps_0,\ldots,2\eps_0,\eps_0,$
were proportional to the exponential series $\{2_{k-1},2_{k-2},\ldots,4,2,1\}$.
For example, $\FF_2$  on 3~spins results in $\{2\eps_0,\eps_0,\eps_0\}$ 
(see Example~6), and $\FF_3$ on 4~spins results in 
$\{4\eps_0,2\eps_0,\eps_0,\eps_0\}$ (see Example~8). This 
coincides with the $k$-step Fibonacci sequence, where $a_{\ell}= 2^{\ell-2}$, 
for $a_1 = 1$ and $\ell = 2,3,...,k+1$ (see 
eq~\ref{eq:k-step-Fibonacci-sequence}). This property leads to cooling 
of $n+1$ spins by a special case of $k$-bonacci (Algorithm~5), 
where $k$-bonacci is applied to $k = n - 1$ spins. 
\\
\textbf{Algorithm~5: } \emph{All-bonacci: $\AA(A_n,\ldots,A_1,m_n\ldots,m_3)$\\
Apply $\mathcal{F}_{n-1}(A_n,\ldots,A_1,m_n,\ldots,m_3)$.
%\label{alg:nbon}
}\\
The final biases after all-bonacci are $\eps_{A_i}\rightarrow\eps_02^{i-2}$
for $i>1.$
The resulting series is $\{\ldots,16,8,4,2,1,1\}$.

The all-bonacci algorithm potentially constitutes an optimal AC scheme, as 
explained in section~\ref{sec:optimal}.

\subsection{Density Matrices of the Cooled Spin Systems}\label{rho-cooled-spins}
For a system of three spins in a completely mixed state (CMS), the density 
matrix of each spin is 
% 
% Parag: PPA on three spins. 
% 
For a spin system in a completely mixed state, the density matrix of each spin 
is:
\begin{equation} 
\frac{1}{2}\II=\frac{1}{2} 
\left( 
\begin{array}{cc} 
1  &\\ 
&  1 
\end{array} 
\right). 
\end{equation}
The density matrix of the entire system, which is diagonal, is given by the
tensor product
$\rho_{CM}=\frac{1}{2^3}\II\otimes\II\otimes\II$, where
\begin{equation}\label{eq:diagonal-of-completly-mixed}
Diag(\rho_{CM})=2^{-3}(1,1,1,1,1,1,1,1).
\end{equation}
For three spins in a thermal state, the density matrix of each spin is
\begin{equation}
\rho_T^{(1)}=\frac{1}{2}
\left(
\begin{array}{cc}
1+\eps_0  &\\
&  1-\eps_0
\end{array}
\right).
\end{equation}
and the density matrix of the entire system is given by the tensor
product $\rho_{T}=\rho_T^{(1)}\otimes\rho_T^{(1)}\otimes\rho_T^{(1)}$. This
matrix is also diagonal. We write the diagonal elements to leading order in
$\eps_0$:
\begin{equation}\label{eq:diagonal-of-thermal}
Diag(\rho_T)=2^{-3}
(1+3\eps_0,1+\eps_0,1+\eps_0,1+\eps_0,1-\eps_0,1-\eps_0,1-\eps_0,1-3\eps_0).
\end{equation}
Consider now the density matrix after shifting and scaling,
 $\rho'=2^{n}(\rho-2^{-n}\II)/\eps_0$.
For any $Diag(\rho)=(p_1,p_2,p_3,\ldots,p_n)$ the resulting diagonal is
$Diag(\rho')=(p_1',p_2',p_3',\ldots,p_n')$ with $p_j'=2^{n}(p_j-2^{-n})/\eps_0$.
 The diagonal of a shifted and scaled (S\&S) matrix for a completely mixed state
(eq~\ref{eq:diagonal-of-completly-mixed}) is
\begin{equation}\label{eq:shifted-diagonal-of-completly-mixed}
Diag(\rho'_{CM})=(0,0,0,0,0,0,0,0),
\end{equation}
and for a thermal state (eq.~\ref{eq:diagonal-of-thermal})
\begin{equation}\label{eq:shifted-diagonal-of-thermal}
Diag(\rho'_{T})=(3,1,1,1,-1,-1,-1,-3).
\end{equation}
%
%YW-11May: removed ref to appendix
%
%See Appendix~\ref{app:SNS-diagonal-thermal-state} for a derivation of the 
%diagonal of a general ($n$ qubit) thermal state under the assumption $n\eps0\ll 1$. 
In the following discussion we 
assume that any element, $p$, satisfies $p'\eps_0\ll1$. When applied to a 
diagonal matrix with elements of the form 
$p=\frac{1}{2^n}\left(1\pm p'_i\eps_0\right)$, this transformation yields the 
corresponding S\&S elements, $\pm p'i$.

Consider now the application of the Fibonacci to three spins. The resultant bias configuration, $\{2\eps_0,\eps_0,\eps_0\}$ (eq~\ref{eq:Fernandez-asymptotic-result}) is associated with the 
density matrix 
\begin{equation}
 \rho_{Fib}^{(3)}=\frac{1}{2^3}
 \left( 
   \begin{array}{cc}
     1+2\eps_0 & \\
               & 1-2\eps_0
   \end{array}
 \right)
 \otimes
 \left( 
   \begin{array}{cc}
     1+\eps_0 & \\
               & 1-\eps_0
   \end{array}
 \right)
 \otimes
 \left( 
   \begin{array}{cc}
     1+\eps_0 & \\
               & 1-\eps_0
   \end{array}
 \right),
\end{equation}
with a corresponding diagonal, to leading order in $\eps_0$,
\begin{equation}
 Diag\left(\rho_{Fib}^{(3)}\right)=
  2^{-3}\left(1+4\eps_0,1+2\eps_0,1+2\eps_0,1,
              1,1-2\eps_0,1-2\eps_0,1-4\eps_0\right).
\end{equation}
The S\&S form of this diagonal is
\begin{equation}\label{eq:rho'-of-3-spins-lower-limit}
 Diag\left({\rho'}_{Fib}^{(3)}\right)=
  \left(4,2,2,0,0,-2,-2,-4\right).
\end{equation}
Similarly, the S\&S diagonal for Tribonacci on four spins is
\begin{equation}\label{eq:rho'-of-4-spins-lower-limit}
 Diag\left({\rho'}_{Trib}^{(4)}\right)=
  \left(8,6,6,4,4,2,2,0,0,-2,-2,-4,-4,-6,-6,-8\right).
\end{equation}
and the S\&S form of the diagonal for all-bonacci on $n$ spins is
\begin{equation}\label{eq:rho'-of-n-spins-lower-limit}
 Diag\left({\rho'}_{allb}^{(n)}\right)=
  \left[2^{n-1},(2^{n-1}-2),(2^{n-1}-2),\ldots,2,2,0,0,\ldots,-2^n-1\right].
\end{equation}
which are good approximations as long as $2^n\eps_0\ll1.$

\subsubsection{Partner Pairing Algorithm}
Recently a cooling algorithm was devised that achieves a superior bias than 
previous AC algorithms.~\cite{SMW07,SMW05} This algorithm, termed the Partner 
Pairing Algorithm (PPA), was shown to produce the highest possible bias for an 
arbitrary number of spins after any number of reset steps. Let us assume that 
the reset spin is the least significant bit (the rightmost spin in the 
tensor-product density matrix). The PPA on $n$ spins is given in 
Algorithm~6. 
\\
\textbf{Algorithm~6: } \emph{Partner Pairing Algorithm (PPA)\\
%\label{alg:PPA}
Repeat the following, until cooling arbitrarily close to the limit.
\begin{enumerate}
\item RESET -- applied only to a single reset spin.
\item SORT -- A permutation that sorts the $2^n$ diagonal elements of the 
              density matrix by decreasing value, such that $p_0$ is the 
              largest, and $p_{2^{n}-1}$ is the smallest.
\end{enumerate}}
Written in terms of $\eps$, the reset step has the effect of changing the 
traced density matrix of the reset spin to
\begin{equation}\label{eq:rho-at-equilibrium}
\rho_{\eps}=\frac{1}{e^{\eps}+e^{-\eps}}
\left(
  \begin{array}{cc}
    e^{\eps} & \\
       &   e^{-\eps}
  \end{array}
\right) =
\frac{1}{2}
\left(
  \begin{array}{cc}
    1+\eps_0  &\\
    &  1-\eps_0
  \end{array}
\right),
\end{equation}
for any previous state. From eq~\ref{eq:rho-at-equilibrium} it is clear that 
$\eps_0=\tanh\eps$ as stated in the introduction. For a single spin in any 
diagonal mixed state:
\begin{equation}
 \left(
  \begin{array}{cc}
   p_0 & \\
    &  p_1
  \end{array}
 \right) 
 \xrightarrow{RESET}\frac{p_0+p_1}{2}
 \left(
  \begin{array}{cc}
    1+\eps_0  &\\
    &  1-\eps_0
  \end{array}
\right)=\frac{1}{2}
\left(
  \begin{array}{cc}
    1+\eps_0  &\\
    &  1-\eps_0
  \end{array}
\right).
\end{equation}
For two spins in any diagonal state a reset of the least significant bit 
results in
\begin{eqnarray}
 &  & \left(
  \begin{array}{cccc}
   p_0  &   &   &\\
    &  p_1  &   &\\
    &   &  p_2  &\\
    &   &   &  p_3
  \end{array}
 \right) 
 \xrightarrow{RESET} \nonumber \\
 &  & \frac{p_0+p_1}{2}
 \left(
  \begin{array}{cccc}
    \frac{p_0+p_1}{2}(1+\eps_0)   &           &           &\\
                & \frac{p_0+p_1}{2}(1-\eps_0) &           &\\
                &           & \frac{p_2+p_3}{2}(1+\eps_0) &\\
                &           &           & \frac{p_2+p_3}{2}(1-\eps_0)
  \end{array}
 \right).
\end{eqnarray}
Algorithm~6 may be highly inefficient in terms of logic gates. Each SORT could 
require an exponential number of gates. Furthermore, even 
calculation of the required gates might be exponentially hard.

We refer only to diagonal matrices and the diagonal elements of the matrices 
(applications of the gates considered here to a diagonal density matrix do not 
produce off-diagonal elements). For a many-spin system, RESET of the reset 
spin, transforms the diagonal of any diagonal density matrix, $\rho$, as 
follows: 
\begin{eqnarray}\label{eq:reset-in-PPA}  
&&\hskip -3ex Diag(\rho)=(p_0,p_1,p_2,p_3,\ldots)\rightarrow \\
&& \left[\frac{p_0+p_1}{2}(1+\eps_0), 
\frac{p_0+p_1}{2}(1-\eps_0), \frac{p_2+p_3}{2}(1+\eps_0), 
\frac{p_2+p_3}{2}(1-\eps_0), \ldots\right], \nonumber 
\end{eqnarray}
as the density matrix of each pair, $p_i$ and $p_{i+1}$ (for even $i$) is 
transformed by the RESET step as described by eq~\ref{eq:rho-at-equilibrium} 
above. We use the definition of S\&S probabilities, $p' = 2^n(p-2{-n})/\eps_0$. The resulting S\&S diagonal is
\begin{eqnarray} 
&&\hskip -3ex Diag(\rho')=\nonumber\\
&&\left[\frac{2^n}{\eps_0}\left(\frac{p_0+p_1}{2}+(1+\eps_0)-2^{-n}\right),
  \ldots \right]= \nonumber\\
&&\left[\frac{2^n}{\eps_0}\left(\frac{p_0+p_1}{2}-2^{-n}\right)
   +2^n\frac{p_0+p_1}{2},\ldots \right]= \nonumber\\
&&\left[\frac{p'_1+p'_2}{2}+1, 
 \frac{p'_0+p'_1}{2}+2^n\frac{p_0+p_1}{2},  
 \frac{p'_0+p'_1}{2}-2^n\frac{p_0+p_1}{2}, \ldots\right],
\end{eqnarray}
where the second element is shown in the final expression. We now use 
$p = 2{-n}(\eps_0 p'+1)$, to obtain
\begin{eqnarray}
2^n\frac{p_0+p_1}{2}&=&2^n\frac{2^{-n}(\eps_0p'_0+1)+2^{-n}(\eps_0p'_1+1)}{2}\\
                    &=&\frac{\eps_0(p'_0+p'_1)+2}{2}
                     = 1+\eps_0\frac{p'_0+p'_1}{2}.
\end{eqnarray}
Ref~\cite{SMW07} provides an analysis of the PPA\@. We continue, as in the
previous subsection, to analyze the case of $p'\eps_0\ll1$, which is of 
practical interest. In this case $1+\eps_0\frac{p'_0+p'_1}{2}\approx 1$. Hence, 
the effect of a RESET step on the S\&S diagonal is:
\begin{eqnarray}\label{eq:RESET-on-shifted-diagonal} 
  Diag(\rho')&=&(p'_0,p'_1,p'_2,p'_3,\ldots)\rightarrow\\
&&\left[\frac{p'_0+p'_1}{2}+1, 
 \frac{p'_0+p'_1}{2}-1, \frac{p'_2+p'_3}{2}+1, \frac{p'_2+p'_3}{2}-1, 
\ldots\right] \nonumber
\end{eqnarray}
Consider now three spins, such that the one at the right is a reset spin. 
Following ref~\cite{SMW07}, we initially apply the PPA to three spins which are 
initially at the completely mixed state. The first step of the PPA, RESET 
(eq~\ref{eq:RESET-on-shifted-diagonal}), is applied to the diagonal of the 
completely mixed state (eq~\ref{eq:shifted-diagonal-of-completly-mixed}), to 
yield 
\begin{equation}\label{eq:Diag-reset} 
Diag(\rho_{CMS}')\rightarrow Diag(\rho'_{RESET})=(1,-1,1,-1,1,-1,1,-1). 
\end{equation}
This diagonal corresponds to the density matrix 
\begin{equation} 
\rho_{RESET}=\frac{1}{2^3} 
\II\otimes\II\otimes\left( 
\begin{array}{cc} 
1+\eps_0 &\\ 
& 1-\eps_0 
\end{array} 
\right), 
\end{equation}
namely to the bias configuration $\{0,0,\eps_0\}.$ The next PPA step, SORT, 
sorts the diagonal elements in decreasing order: 
\begin{equation} 
Diag(\rho'_{SORT})=(1,1,1,1,-1,-1,-1,-1), 
\end{equation}
that arises from the density matrix 
\begin{equation} 
\rho_{SORT}=\frac{1}{2^3} 
\left( 
\begin{array}{cc} 
1+\eps_0 &\\ 
& 1-\eps_0 
\end{array} 
\right) 
\otimes\II\otimes\II, 
\end{equation}
which corresponds to the biases $\{\eps_0,0,0\}$. The bias was thus 
transferred to the leftmost spin. In the course of repeated alternation 
between the two steps of the PPA, this diagonal will further evolve as 
detailed in Example~12.

The rightmost column of Example~12 lists the resulting 
bias configurations. Notice that after the 5$^{\mathrm{th}}$ step the biases 
are identical. Also notice that in the 6$^{\mathrm{th}}$ and 10$^{\mathrm{th}}$ 
steps the states \ket{100} and \ket{011} are exchanged (3B-Comp); after both 
these steps, the state of the system cannot be written as a tensor 
product.\footnote{This is due to \emph{classical} correlations (not involving 
entanglement) between the spins; an individual bias may still be associated 
with each spin by tracing out the others, but such a bias cannot be interpreted 
as temperature.} Also notice that after step~13, the PPA applies a SORT which 
will also switch between these two states. The PPA, applied to the bias 
configuration $\{t\eps_0,\eps_0,\eps_0\}$, where $1\le t \le 2$ is simply an 
exchange $\ket{100}\leftrightarrow\ket{011}.$ This is evident from the diagonal 
\begin{equation} 
Diag(\rho')=(t+2,t,t,t-2,-t+2,-t,-t,-t-2). 
\end{equation}
Thus, the PPA on three spins is identical to the Fibonacci algorithm applied 
to three spins (see Example~6). This analogy may be taken 
further to some extent. When the PPA is applied onto four spins, the outcome, 
but not the steps, is identical to the result obtained by Tribonacci algorithm 
(see Example~8). These identical results may be generalized to the 
PPA and all-bonacci applied onto $n \ge3$ spins.
\\
\textbf{Example~12: }\emph{Application of PPA to $3$ spins
\def\s{\mathrm{~step~}}
\def\xr{\xrightarrow{RESET}}
\def\xs{\xrightarrow{SORT}}
 \begin{center}
$\begin{array}{l|ll|c} 
  &Diag(\rho')=&(0,0,0,0,0,0,0,0)&\{0,0,0\}\\ 
\s 1&\xr&(1,-1,1,-1,1,-1,1,-1)&\{0,0,\eps_0\}                                \\ 
\s 2&\xs&(1,1,1,1,-1,-1,-1,-1)&\{\eps_0,0,0\}                                \\ 
\s 3&\xr&(2,0,2,0,0,-2,0,-2)  &\{\eps_0,0,\eps_0\}                           \\ 
\s 4&\xs&(2,2,0,0,0,0,-2,-2)  &\{\eps_0,\eps_0,0\}                           \\ 
\s 5&\xr&(3,1,1,-1,1,-1,-1,-3)&\{\eps_0,\eps_0,\eps_0\}                      \\ 
\s 6&\xs&(3,1,1,1,-1,-1,-1,-3)& N.~A.                                        \\ 
\s 7&\xr&(3,1,2,0,0,-2,-1,-3) &\{\frac{3\eps_0}{2},\frac{\eps_0}{2},\eps_0\} \\ 
\s 8&\xs&(3,2,1,0,0,-1,-2,-3) &\{\frac{3\eps_0}{2},\eps_0,\frac{\eps_0}{2}\} \\ 
\s 9&\xr&\frac{1}{2}(7,3,3,-1,1,-3,-3,-7) 
                              &\{\frac{3\eps_0}{2},\eps_0,\eps_0\}           \\ 
\s 10&\xs&\frac{1}{2}(7,3,3,1,-1,-3,-3,-7) & N.A.                            \\
\s 11&\xr&\frac{1}{2}(7,3,4,0,0,-4,-3,-7)
                              &\{\frac{7\eps_0}{4},\frac{3\eps_0}{4},\eps_0\}\\
\s 12&\xs&\frac{1}{2}(7,4,3,0,0,-3,-4,-7)~~~~~~\strut
                              &\{\frac{7\eps_0}{4},\eps_0,\frac{3\eps_0}{4}\}\\
\s 13&\xr&\frac{1}{4}(15,7,7,-1,1,-7,-7,-15)
                              &\{\frac{7\eps_0}{4},\eps_0,\eps_0\} 
\end{array}$
\end{center}
%\label{ex:PPA-on-3-spins}
}

\section{Optimal Algorithmic Cooling}\label{sec:optimal}
\subsection{Lower Limits of the PPA and All-bonacci}
Consider an application of the PPA to a three-spin system at a state with the 
S\&S diagonal of eq~\ref{eq:rho'-of-3-spins-lower-limit}. This diagonal is 
both sorted and invariant to RESET, hence it is invariant to the PPA\@. 
Eq~\ref{eq:rho'-of-3-spins-lower-limit} corresponds to the bias configuration 
$\{2\eps_0, \eps_0, \eps_0\}$ which is the limit of the Fibonacci algorithm 
presented above (eq~\ref{eq:Fernandez-asymptotic-result}). This configuration 
is the ``lower'' limit of the PPA with three spins in the following sense: any 
``hotter configuration'' is not invariant and continues to cool down during 
exhaustive PPA until reaching or bypassing this configuration. For four spins 
the diagonal of Eq~\ref{eq:rho'-of-4-spins-lower-limit} is invariant to the PPA
for the same reasons. This diagonal corresponds to the bias configuration 
$\{4\eps_0, 2\eps_0, \eps_0, \eps_0\}$, which is the limit of the 
Tribonacci algorithm (or the all-bonacci) applied to four spins. Any ``hotter 
configuration'' is not invariant and cools further during exhaustive PPA, until 
it reaches (or bypasses) this configuration. 

Now, we follow the approximation of very small biases for $n$ spins, where 
$2^n\eps_0\ll 1$. The diagonal in eq~\ref{eq:rho'-of-n-spins-lower-limit} is 
invariant to the PPA\@. This diagonal corresponds to the bias configuration
\begin{equation}\label{eq:biases-lower-limit}
 \{2^{n-2}\eps_0,2^{n-3}\eps_0,2^{n-4}\eps_0,\ldots,2\eps_0,\eps_0,\eps_0\},
\end{equation}
which is the limit of the all-bonacci algorithm. As before, any ``hotter 
configuration'' is not invariant and cools further. It is thus proven that the 
PPA reaches at least the same biases as all-bonacci.

We conclude that under the assumption $2^n\eps_0\ll1$, the $n$-spin system can
be cooled to the bias configuration shown in eq~\ref{eq:biases-lower-limit}. 
When this assumption is not valid (e.g., for larger $n$ with the same $\eps_0$), pure qubits can be extracted; see theorem 2 in~\cite{SMW05} 
(theorem 3 in ref~\cite{SMW07}).

Yet, the PPA potentially yields ``colder configurations''. We obtained numerical 
results for small spin systems which indicate that the limits of the PPA and 
all-bonacci are identical. Still, since the PPA may provide better cooling, it 
is important to put an ``upper'' limit on its cooling capacity.

\subsection{Upper Limits  on Algorithmic Cooling}
Theoretical limits for cooling with algorithmic cooling devices have recently 
been established~\cite{SMW07,SMW05}. For any number of reset steps, the PPA has
been shown to be optimal in terms of entropy extraction (see ref~\cite{SMW05} 
and more details in section~1 of ref~\cite{SMW07}). An upper bound on the 
degree of cooling attainable by the PPA is therefore also an upper bound of any 
AC algorithm. The following theorem regards a bound on AC which is the loose 
bound of ref~\cite{SMW07}.
\begin{thm}
\label{thm:upper-bound} No algorithmic cooling
method can increase the probability of any basis state to 
above\footnote{A tighter bound, 
$p_0\le2^{-n}e^{2^{n-1}\eps}$, was claimed by theorem~1 of ref~\cite{SMW07}.} 
  $\min\{ 2^{-n}e^{2^{n}\eps},1\}$,
where the initial configuration is the completely mixed state.\footnote{It is 
assumed that the computation spins are initialized by the polarization of the 
reset spins.} This includes the idealization where an unbounded number of reset 
and logic steps can be applied without error or decoherence.
\end{thm}
The proof of Theorem~\ref{thm:upper-bound} involves applying the PPA and showing
that the probability of any state never exceeds $2^{-n}e^{2^{n}\eps}.$  

\section{Algorithmic Cooling and NMR Quantum Computing}\label{sec:NMR-QC}
We have been using the language of classical bits, however spins are quantum 
systems; thus, spin particles (two-level systems) should be regarded as 
quantum bits (qubits). A molecule with $n$ spin nuclei can represent an 
$n$-qubit computing device. The quantum computing device in this case is 
actually an ensemble of many such molecules. In ensemble NMR quantum 
computing~\cite{CFH96,CFH97,GC97} each computer is represented by a single 
molecule, such as the TCE molecule of Fig.~\ref{fig:TCE}, and the qubits of 
the computer are represented by nuclear spins. The macroscopic number of 
identical molecules available in a bulk system is equivalent to many 
processing units which perform the same computation in parallel. The molecular 
ensemble is placed in a constant magnetic field, so that a small majority of 
the spins are aligned with the direction of the field. To perform a desired 
computation, the same sequence of external pulses is applied to all 
molecules/computers. Any quantum logic-gate can be implemented in NMR by a 
sequence of radio-frequency pulses and intermittent delay periods during which 
spins evolve under coupling~\cite{PHC00}. Finally, the state of a particular 
qubit is measured by summing over all computers/molecules. The process of AC 
constitutes a simple quantum computing algorithm. However, unlike other 
quantum algorithms, the use of quantum logic gates does not produce any 
computational speed-up, but instead generates colder spins. This constitutes 
the first near-future application of quantum computing devices. AC may also 
have an important long-term application; it may enable quantum computing 
devices capable of running important quantum algorithms, such as the 
factorization of large numbers~\cite{Shor97}. NMR quantum 
computers~\cite{CFH96,CFH97,GC97} are currently the most successful quantum 
computing devices (see for instance ref~\cite{VSB+01}), but are known to 
suffer from severe scalability problems~\cite{BMRVV02,Warren97,DiVincenzo98}. 
AC can be used for building scalable NMR quantum computers of 20-50 quantum 
bits if electron spins are used for the PT and RESET steps. PT with electron 
spins~\cite{FHG+01,AFGH+03} can enhance the polarization by three or four 
orders of magnitude. Unfortunately, severe technical difficulties have thus 
far impeded common practice of this technique. An example of such a difficulty 
is the need to master two very different electromagnetic frequencies within a 
single machine. However, in case such PT steps come into practice (using 
machinery that allows conventional NMR techniques as well), AC could be 
applied with much better parameters; First, $\eps_0$ could be increased to 
around 0.01-0.1. Second, the ratio $R_{relax-times}$ could reach $10^3-10^4$. 
With these figures, scalable quantum computers of 20-50 qubits may become 
feasible.

\section{Discussion}
Algorithmic Cooling (AC) harnesses the environment to enhance spin 
polarization much beyond the limits of reversible polarization compression 
(RPC). Both cooling methods may be regarded as a set of logic gates, such as 
NOT or SWAP, which are applied onto the spins. Polarization transfer (PT), 
for instance, a form of RPC, may be obtained by a SWAP gate. AC algorithms are 
composed of two types of steps: reversible AC steps (RPC) applied to two spins 
or more, and reset steps, in which entropy is shifted to the environment 
through reset spins. These reset spins thermalize much faster than computation 
spins (which are cooled), allowing a form of a molecular heat pump. A 
prerequisite for AC is thus the mutual existence (on the same molecule) of two 
types of spins with a substantial relaxation times ratio (of at least an order 
of magnitude). While this demand limits the applicability of AC, it is often 
met in practice (e.g., by \isotope{H}{1} vs \isotope{C}{13} in certain organic 
molecules) and may be induced by the addition of a paramagnetic reagent. The 
attractive possibility of using rapidly-thermalizing electron spins as reset 
bits is gradually becoming a relevant option for the far future. 

We have surveyed previous cooling algorithms and suggested novel algorithms 
for exhaustive AC: the Tribonacci, the 
$k$-bonacci, and the all-bonacci algorithms. We conjectured the optimality of 
all-bonacci, as it appears to yield the same cooling level as the PPA of 
refs~\cite{SMW07,SMW05}. Improving the SNR of NMR by AC potentially 
constitutes the first short-term application of quantum computing devices. AC 
is further accommodated in quantum computing schemes (NMR-based or others) 
relating to the more distant future~\cite{BMRVV02,Twamley03,FS03,LGYY+02}. 

\section{Acknowledgements.} Y.E., T.M., and Y.W. thank the Israeli Ministry of 
Defense, the Promotion of Research at the Technion, and the Institute for 
Future Defense Research for supporting this research.
\bibliographystyle{splncs}
\bibliography{/home/yossiv/qcomp/mine}
\end{document}